\documentclass[12pt, a4paper]{article}
\usepackage{amsmath, amsthm,  amsfonts}

\title{\bf{A note on the  $(1, 1, \dots, 1)$ monopole metric.}}
\author{Michael K. Murray\footnote{
Pure Mathematics Department,
University of Adelaide,
Australia, 5005,
email: mmurray@maths.adelaide.edu.au}}

\def\R{{\mathbb R}}
\def\Z{{\mathbb Z}}

\def\H{{\mathbb H}}

\def\A{{\cal A}}

\def\G{{\cal G}}
\def\N{{\cal N}}

\def\cm{\text{c}}

\def\Re{\text{Re}}
\def\Im{\text{Im}}

\def\<{\langle}
\def\>{\rangle}

\numberwithin{equation}{section}

\theoremstyle{plain}

\theoremstyle{definition}

\theoremstyle{remark}

\begin{document}
\maketitle

\section{Introduction}
For some time now there has been considerable interest
in the natural hyperkahler metric on the moduli space
of charge $m$ $SU(2)$ monopoles in $\R^3$.
It is know from the work of Taubes that near
the boundary of this moduli space
the monopole approximates a collection of $m$
particles with internal $U(1)$ phases.
It was argued by Manton  \cite{Man}
that the geodesics of this metric
correspond to   scattering of $m$ slow moving monopoles.
There are now many interesting examples of scattering of
$SU(2)$ monopoles beginning
with the calculation of the metric on the moduli space of
$SU(2)$ charge two
monopoles  by Atiyah and Hitchin  \cite{AtiHit} and more recently
results on the scattering of monopoles with special symmetry
\cite{HitManMur, HouSut, HouSut1, HouSut2, HouSut3}.

Monopoles also exist for compact groups $G$ other than $SU(2)$. We
will be interested only in the case of maximal symmetry breaking.
In this case the particles making up the monopole come in
$r$ distinguishable `types' where $r$ is the rank of the group.
The $r$ types correspond correspond to the the $r$
different elementary ways of embedding $SU(2)$ into $G$ along
simple root directions.
The magnetic charge of the monopole is now a vector
$m = (m_1, \dots, m_r)$ where  $m_i$ can be thought of as the
number of monopoles of type $i$ \cite{Mur}.
If any of the $m_i$ vanish the monopole is obtained from
an embedded subgroup so that the simplest monopole that is genuinely a
monopole for $G$ is one with each $m_i = 1$.  We are interested in the
structure of the moduli space for this case and its metric. Note
that in general the moduli space has dimension $4(m_1+ \dots + m_r)$
so that  the moduli space of $(1, 1, \dots, 1)$ monopoles
has dimension $4r$.

For the group $SU(3)$  the rank is $2$ and the metric on the moduli
space of $(1,1)$ monopoles was studied by Connell
\cite{Con, Con1}.
The same result was also obtained independently Gauntlett
and Low  \cite{GauLow} and Lee, Weinberg and  Yi
\cite{GauLow, LeeWeiPil}. In these latter two works
some special assumptions on the  values of the Higg's field at
infinity that simplified the
work of  Connell are removed. The metric obtained
is globally of Taub-NUT type.

For the more general
case of an $SU(n+1)$ monopole of charge $(1, \dots, 1)$ Lee, Weinberg
and Yi
\cite{LeeWeiPil1} calculate the asymptotic form of the monopole
metric and show that it is asymptotically Taub-NUT. They
then give an argument that the asymptotic form of the metric
can be smoothly extended to the whole moduli space and they
conjecture that the monopole metric is indeed exactly this
extended metric.  I give a partial proof of this result.
The reason it is partial is that I construct and
describe the natural hyperkaehler metric not on the
monopole moduli space but on the space of {\em Nahm data}.
This is indeed of the form conjectured in \cite{LeeWeiPil1}.
Moreover it is known \cite{Nah, HurMur} that the moduli space
of Nahm data is diffeomorphic to the moduli space of monopoles.
In the case of $SU(2)$ is also known that this diffeomorphism is an
isometry \cite{Nak} but for other $SU(n+1)$ groups, while
this is believed to be true, it has not yet been proved.

In summary the paper is as follows: In Section \ref{sec-HKq} I
review the hyperkaehler quotient construction applied
to quaternionic vector spaces.  In Section  \ref{sec-infHKq} I
describe the infinite dimensional hyperkaehler quotient
that defines $\N$ the moduli space of $(1, \dots, 1)$
Nahm data and show it can be realised as a finite-dimensional
hyperkahler quotient. This enables a rigourous definition
of the metric on $\N$ as a hyperkaehler quotient of a finite
dimensional hyperkaehler manifold. This is described in
Section \ref{sec-finHKq} and in Section \ref{sec-mod} it is
shown that the moduli space is isometric to a product
$$
\N = \frac{\N \times \R^3 \times \R}{\Z}
$$
where $\N_c$ is the space of  centered Nahm
data corresponding to strongly centered monopoles and
$\R^3 \times \R$ is given a  multiple of the standard
metric. Finally in Section \ref{sec-metric} we
consider the metric on $\N_c$. The space $\N_c$
is just $\H^{n-2}$ where $\H = \R^4$ is
quaternionic space. In the case of $SU(3)$
it is possible to give an explicit formula for the metric
on this space \cite{Con, Con1}, in the present case I
use a result of Hitchin \cite{Hit} to show that
it has the same form as the
metric in \cite{LeeWeiPil1}.

\section{Hyperkaehler quotients of  vector spaces.}
\label{sec-HKq}
A hyperkaehler manifold \cite{Hit} is a
Riemannian manifold $(M, g)$ with three  complex structures
$I$, $J$ and $K$ which satisfy
the  quaternion algebra and are covariantly constant.

We need to consider  from \cite{HitKarLinRoc} the hyperkaehler quotient
of a hyperkaehler manifold by a group.
For our purposes it is enough to consider the case
when the manifold that is being quotiented is a
vector space.
Let
$V$ be a real vector space with three complex structures
$e_1 = I$, $e_2 = J$, $e_3 = K$ which satisfy
the quaternion algbra. Assume also that $V$ has an inner
product $\<\ ,\ \>$ which is preserved by each of the $e_i$.
Then $V$ has three symplectic forms $\omega_i$ defined by
$\omega_i(v, w) = \<v, e_i w\>$. Because the tangent
space at any point of $V$ is canonically identified with
$V$ itself this makes $V$ a hyperkaehler manifold.

Assume now that a group $G$ acts freely on $V$ in
such a way that $V/G$ is a manifold and $V \to V/G$ is a
principal $G$ fibration. Assume further that the $G$ action
preserves the inner product on the tangent spaces of $V$
and also commutes with the action of the $e_i$. If $\xi$ is
an element of $LG$, the Lie algebra of $G$, it defines a
vector field $\iota(\xi) $ on $V$. The  {\em moment map}
$$
\mu \colon V \to \R^3 \otimes LG^*
$$
of this group action is then defined by
\begin{align}
\mu_k(v) &= \int_0^1 \omega_k(\iota(\xi), v) dt \notag \\
   &= \int_0^1 \< \iota(\xi), e_k(v)\> dt
\label{eq:moment}
\end{align}

The hyperkaehler quotient  of $V$ is the space $\mu^{-1}(0)/G$.
To see that this is  a hyperkaehler manifold
let $\pi$ be the projection $\pi\colon \mu^{-1}(0) \to \mu^{-1}(0)/G$.
If $x \in \mu^{-1}(0)/G $ choose $\hat x \in \pi^{-1}(x)$.
We can split the tangent
space at $\hat x$ into vertical directions tangent to the $G$ action
and horizontal directions which are orthogonal to the vertical
directions. The horizontal directions are naturally identified
with the tangent space to $\mu^{-1}(0)/G $ at $x$ and this
enables us to define an inner product and a hyperkaehler
structure on that tangent space. This construction is, in fact,
independent of the choice of $\hat x$ in $\pi^{-1}(\pi(x)$
because of the $G$ invariance.  I refer
the reader to \cite{HitKarLinRoc} for details.

\section{The moduli space of Nahm data.}
\label{sec-infHKq}

We are interested in $SU(n+1)$ monopoles or more precisely
their Nahm data. In the interests of brevity I will not review
the theory of monopoles
or the relationship between monopoles and solutions
of Nahm's equations but refer the reader to \cite{AtiHit}
and references therein for the $SU(2)$ theory and to
\cite{Mur, HurMur, Nah} for the $SU(n+1)$ theory.
We will denote by $\N$ the moduli space of Nahm data
which is realised as follows. The Higg's field at infinity
of the monopole has eigenvalues
$i\lambda_n, \dots, i\lambda_0$ where we assume that
$$
\lambda_n< \lambda_{n-1} < \dots < \lambda_1 < \lambda_0.
$$
In the monopole language this means we have maximal
symmetry breaking at infinity.

Denote by $\A $ the set of all pairs $(T, a) $ where $ a \in \H^{n-1}$ and
$T \colon [\lambda_n, \lambda_0] \to \H$ with the property
that the restriction of $T$ to each interval $(\lambda_i , \lambda_{i-1})$
is smooth and has a smooth extension to $[\lambda_i , \lambda_{i-1}]$.
We denote this extension by
$$
T^i \colon [\lambda_i , \lambda_{i-1}] \to \H.
$$
The map $T$ itself is allowed to have  discontinuities at the $\lambda_i$.
It is useful to think of the vector
$a= (a^1, \dots, a^{n-1}$ as a function on the set
$(\lambda_n, \dots, \lambda_0)$ whose
value at $\lambda_i$ is just $a^i$.
We will consider the space $\A$ as a left quaternionic vector space.

Denote by $\G$ the group of all continous maps $g \colon [\lambda_n,
\lambda_0] \to U(1)$ which are smooth on an open subinterval $(\lambda_i ,
\lambda_{i-1})$ and whose derivatives may be discontinous at the points
$\lambda_i$ for $i = 1, \dots, n-1$ but such that the restriction of $g$ to
$(\lambda_i , \lambda_{i-1})$ has a smooth extension to $[\lambda_i ,
\lambda_{i-1}]$. We require further that $g(\lambda_n) = g(\lambda_0) = 1$.
We denote by $g^i$ the extension to $[\lambda_i , \lambda_{i-1}]$ of the
restriction of $g$ to $(\lambda_i , \lambda_{i-1})$.

The group $\G$ acts on the left of $\A$ by
\begin{align*}
(gT)^j &= T^j + \frac{1}{i}  \frac{dg^j}{g^j}\cr
(ga)^j &= a^j g(\lambda^j). \cr
\end{align*}
Notice that by continuity $g(\lambda_i) =
g^j(\lambda_j)= g^{j+1}(\lambda_j)$.

We define an inner product on $\A$ by
$$
\< (T, a), (S, b)\> = \sum_{i=1}^n \int_{\lambda_{i-1}}
^{\lambda_i} \Re(T^i\bar S^i)
+ \sum_{i=1}^{n-1} \Re(a^i \bar b^i).
$$

This inner product makes $\A$ an (infinite-dimensional) hyperkaehler
vector space.   We want to consider its hyperkaehler quotient.
It is easy to check that the group action preserves the hyperkaehler
structure. It is not clear, because of the infinite dimensionality,
that the quotient is nicely behaved. We will avoid confronting
this problem by showing that we can replace $\A$ by a finite
dimensional vector space and form the hyperkaehler quotient of
that instead.

To define the moment maps for the action of $\G$ we need to consider the
infinitesimal action of the Lie algebra $L\G$. The Lie algebra $L\G$ is the
set of all continous maps $\xi \colon [\lambda_n, \lambda_0] \to R$ with
$\xi(\lambda_n) = \xi(\lambda_0) = 0$ and whose derivative may jump at
$\lambda_i$ for $i = 1, \dots, n-1$. The
derivative has a smooth extension from $(\lambda_i, \lambda_{i-1})$ to
$[\lambda_i, \lambda_{i-1}]$ and we denote this smooth extension by
$\xi^i$.  We fix our conventions by defining the exponentional map for the
group $\G$ to be $ \xi \mapsto \exp(2\pi i \xi)$. The element $\xi \in L\G$
then defines a vector field $\iota(\xi)$ on $\A$ whose value at $(T, a)$
is
$$
(\iota(\xi)(T, a))^j = (2\pi d\xi^j, 2\pi a^j i \xi^j).
$$
Here and below we sometimes adopt the notation
$$
X^j = (T^j, a^j)
$$
to mean
$$
X = ((T^1, \dots, T^n), (a^1, \dots, a^{n-1})).
$$

We can now calculate the moment map from \eqref{eq:moment}
and we find that $(T, a)$ is in the kernel of $\mu$ if and
only if
$$
\Re(dT^j) = 0
$$
for each $j= 1, \dots, n$ and
$$
\Im(T^{j+1} - T^j) = \frac{1}{2} a^j i\bar a^j
$$
for each $j = 1, \dots, n-1$.

It is clear from these equations that to describe the hyperkaehler
quotient $\N$ of $\A$ by $\G$ we could restrict our attention
from $\A$ to the subset of pairs $(T, a)$ where the imaginary
part of $T$ is constant. If we do that and wish to still have a
hyperkaehler structure then we will need to restrict attention
to $T$ whose real part is also constant.  Notice that if we start
out with a $T$ which is real then by integrating starting at
$\lambda_n$ we can construct a $g \colon [\lambda_n, \lambda_0]
\to U(1)$ such that $gT = 0$ and satisfying all the conditions
to be in $\G$ except that we may not have $g(\lambda_0) = 1$.
But in that case we can find an $h$ such that $dh$ is
constant and $h(\lambda_0) = g(\lambda_0)^{-1}$. The composite
$gh$ is in $\G$ and $gT$ has constant real part. We conclude
that every $(T, a)$  in $\mu^{-1}(0)$ can be gauge transformed
so that $gT$ is constant.

\section{The hyperkaehler quotient.}
\label{sec-finHKq}
Denote by $\A_0$ the set of all triples $(\tau, x, a) $
where $\tau \in \R^n$, $x \in \Im(\H)^n$ and $a  \in \H^{n-1}$.
We identify $\A_0$ with a subset of $\A$ by mapping
each $(\tau, x, a)$ to $(T = \tau + x, a)$ where we
think of $T$ as a step function on $[\lambda_n, \lambda_0]$
whose value on $[\lambda_i, \lambda_{i-1}]$ is $T^i = \tau^i + x^i$.
We shall identify $x^i \in \Im(\H)$ with the corresponding element of
$\R^3$ and call it  the location of the $i$th monopole. It follows
from the discussion at the end of Section \ref{sec-HKq} that the
hyperkaehler quotient of $\A_0$ by $\G_0$ is the same
as the hyperkaehler quotient of $\A $ by $\G$ and hence
yields $\N$ the moduli space of Nahm data.

The space $\A_0$ is a quaternionic vector space and has an inner
product induced from $\A$ which is
\begin{equation}
\<(\tau, x, a), (\sigma, y, b )\>
= \sum_{i=1}^n p_i \tau^i\sigma^i + \sum_{i=1}^n p_i \Re(x^i\bar y^i) +
 \sum_{i=1}^n \Re(a^i\bar b^i)\label{eq:ip}
\end{equation}
where $p_i = \lambda_i - \lambda_{i-1}$.

Consider the subgroup $\G_0 \subset \G$ that fixes $\A_0$. This
is the group of all $g \in \G$ such that  $dg$ is a step function
on $[\lambda_n, \lambda_0]$. That is each $dg^i $ is a constant.
Such a $g$ can be written as
$$
g^j(s) = \exp\bigl(\frac{2i\pi}{p_j}((W^{j}_+ - W^{j-1}_-)s +
                      W^{j-1}_- \lambda_j - W^j_+ \lambda_{j-1})\bigr).
$$
Notice that $g^j(\lambda_j) = \exp(2\pi i W^j_-)$ and $g^j(\lambda_{j-1}) =
\exp(2\pi i W^{i-1}_+)$ so that the condition for $g$ to be continous is
that $W^j_+ - W^j_- $ is an integer.
The numbers $W^j_-$ and $W^j_+$  are not
uniquely determined by $g$. They can be changed by adding to both of them
the same integer.

The group $\G_0$ acts on $\A_0$ by
$$
g(\tau, x, a) = (g\tau, x, ga)
$$
where
\begin{align*}
(g\tau)^i =& \tau^i + \frac{2\pi}{p_i}(W^i_+ - W^{i-1}_-)\cr
(ga)^i  =& a^i \exp(2\pi i W^i_+) = a^i \exp(2\pi i W^i_-) \cr
\end{align*}
If $\xi \in L\G$ then the vector field $\iota(\xi)$
it defines on  $\A_0$  is
$$
(\iota(\xi)(\tau, x, a))^j = (\frac{2\pi}{p_j}(\xi^j
- \xi^{j-1}), 2\pi a^j i \xi^j, 0).
$$

The moment map $\mu$ for the action of $\G_0$ on $\A_0$ can be calculated
from \eqref{eq:moment} but it is the restriction of that for $\G$ on $\A$
and hence we deduce that $(\tau, x, A) \in \mu^{-1}(0)$ if and only if
$$
x^{j+1} - x^j = a^j i \bar a^j
$$
for each $j = 1, \dots, n-1$.

Let $\hat \N = \mu^{-1}(0)$ so the moduli space of
Nahm data is  $\N = \hat \N /\G_0$.

\section{The metric on monopoles.}
\label{sec-mod}
By the Nahm transform \cite{Nah, HurMur} the space $\N$
is diffeomorphic to the space of monopoles of type
$(1, 1, \dots, 1)$. The monopole corresponding
to the orbit of $(\tau, x, a)$ can be interpreted
as a collection of $n$ particles,  located at each of the points $x^j$
with phases $\exp(ip_j\tau^j)$.
Following \cite{LeeWeiPil}
we define the {\em center} of $\tau$ and $x$ by
$$
\tau_{\cm} = \frac {\sum_{i=1}^n p_i \tau^i }{p}
$$
and
$$
x_{\cm} = \frac {\sum_{i=1}^n p_i x^i }{p}.
$$
where
$$
p = \sum_{i=1}^n p_i.
$$
Define $\hat\N_c$ to be the subset of $\hat \N$ consisting of
those $(\tau, x, a)$
with $\tau_{\cm} = 0$ and $x_{\cm} = 0$.
We call this the space of {\em  centered} monopoles.
Define also
$$
\G_{\cm} = \{ g \mid \sum_{i=1}^n W^i_+ - W^{i-1}_- = 0\}.
$$
This is the subgroup of $\G$ which fixes $\hat \N_c$.
We define $\N_c = \hat \N_c/\G_{\cm}$.

We want to define an isomorphism:
\begin{equation}
\hat\N/\G_{\cm} \to \N_c \times \R^3 \times \R. \label{eq:iso}
\end{equation}

To construct the isomorphism  we first define for any  $x \in \R^3$ and
$\tau \in \R$  $\hat x \in (\R^3)^{n-1}$
and $\hat \tau \in \R^{n-1}$ by
$\hat x = (x , x , \dots, x) $ and $\hat \tau = (\tau, \tau, \dots, \tau)$.
Notice that $\hat x_{\cm } = x$ and $\tau_{\cm} = \tau$.
So given a monopole  $(\tau, x, a) \in \hat \N$ we can
center it by defining
$(\tau - \hat \tau_{\cm}, x - \hat x_{\cm} , a) \in \hat \N_c$.
The  map in \eqref{eq:iso} is then defined to send
$(\tau, x, a)$  to the pair
$ ((\tau - \tau_{\cm}, x- x_{\cm} , a) ,
(\tau_{\cm}, x_{\cm}))$ consisting of the
corresponding centered monopole and the center of the
monopole.   This map has inverse given by
$((\tau, x, a), (s, y)) \mapsto (\tau + \hat s, x+\hat y, a)$.

The spaces $\hat\N/\G_{\cm}$ and $\N_c$ inherit inner products
by the process described at the end of Section \ref{sec-HKq}.
 It is straightforward to
calculate that the isomorphism \eqref{eq:iso}
is an isometry if we give $\R^3 \times \R$ the
standard metric multiplied by a factor of $p$.

Finally   notice that $\sum_{i=1}^n W^i_+ - W^{i-1}_- =
\sum_{i=1}^{n-1} W^i_+ - W^{i}_-$ is an integer so that $\G /\G_{\cm}$
is isomorphic to $\Z$. We conclude that there is an isometry
$$
\N  =\frac{\N_c \times \R^3 \times \R}{\Z}.
$$

\section{The metric on centered monopoles.}
\label{sec-metric}
If $(\tau, x, a)$ is in $\N_c$ then the vector $x$
is determined by the equations $x^{j+1} - x^j = a^j i \bar a^j$. So the
triple $(\tau, x, a)$ is determined by the pair $(\tau, a)$. It is
straightforward to show that the orbit of $(\tau, x, a)$ under $\G_{\cm}$
contains exactly one triple of the form $(0, x', a')$. It follows that
$\N_c$ has the topology of $\H^{n-1}$.

In the case that $n=2$ Connell calculated explicitly the hyperkahler
quotient  metric on $\H$. In the case at hand that calculation is more
involved and it is simpler to use an approach due to Hitchin \cite{Hit}.
The $n-1$ dimensional torus $T^{n-1} = U(1)^{n-1}$ acts on the space
$\N_c$ preserving the hyperkaehler metric by rotating each of the $a^i$.
The moment map for the $i$th of these actions is given by $\mu_i(\tau, x, a)
= 2\pi(x^{i+1} - x^i) $ for each $i = 1, \dots, n-1$. This action is free
if none of the $a^j$ vanish. Let $\N'_c$ be the set of $(\tau, x, a)$ such
that noe of the $a^j$ vanish. Denote by $M$ the image of $\N'_c$ in
$(\R^3)^{n-1}$ under the moment map.  The moment map $\N'_c \to M$ realises
$\N_c'$ as a $T^{n-1}$ bundle over $M$. The inner product on $\N_c$ allows
us to define a horizontal subspace orthogonal to the $T^{n-1}$ action at
each point of $\N'_c$ and hence we can define a connection on $\N'_c \to
M$. This defines a  one-form $\alpha = (\alpha^1, \dots, \alpha^{n-1})$
corresponding to projecting onto the vertical subspace. By generalising the
calculation of Hitchin in Section IV.4 of \cite{Hit} it is possible to show
that the metric on $\N'_c$ must  have the form \begin{equation} g =
\sum_{i,j = 1}^{n-1} K_{ij}^{-1} \sum_{a=1}^3 d\mu_i^a d\mu_j^a +
\sum_{i,j=1}^{n-1}  K_{ij} \alpha^i \alpha^j \label{eq:metric}
\end{equation} for some matrix valued function $K_{ij}$ which is constant
in the torus directions.  If $\eta_i$ are the generators of the torus
action then Hitchin's result gives
$$
K_{ij} = g(\eta_i, \eta_j).
$$
We wish to now calculate the $K_{ij}$.

To calculate the $\eta_i$ we have to split them into a vector $\iota(\xi_i)$
in the direction of the $\G_{\cm}$ action and an orthogonal vector
$\hat\eta_i = \eta_i - \iota(\xi_i)$.
Then we have that
$$
K_{ij} = g(\eta_i, \eta_j) = \< \hat\eta_i,\hat \eta_j \>
$$
where $\<\ ,\ \>$ is the inner product defined in
\eqref{eq:ip}. Using the
orthogonality we deduce that
$$
K_{ij} = \< \eta_i, \eta_j\> - \<\iota(\xi_i), \eta_j\>.
$$

The condition that defines the $\iota(\xi_i) $ is the requirement that
$\eta_i - \iota(\xi_i)$ be horizontal, that is:
$$
\<\eta_i - \iota(\xi_i), \iota(\rho)\> = 0
$$
for all $\rho \in L\G_{\cm}$.  Expanding this we have that
$$
\<\eta_i, \iota(\rho)\>  = \<\iota(\xi_i), \iota(\rho)\>.
$$
The vector  $\eta_l$ is
$$
\eta_l( \tau, x, T)^j = (0, 0, \delta_{lj} 2\pi i a^l)
$$
and hence we have
$$
\<\eta_l, \iota(\rho)\> = 2\pi \rho_i |a^l|^2.
$$
The other inner product is
\begin{align}
\< \iota(\xi_l), \iota(\rho)\>&= 4\pi^2 ( \sum_{k=1}^n \frac{1}{p_k}
(\xi_l^k - \xi_l^{k-1})(\rho^k - \rho^k) +
4\pi^2\sum_{k=1}^{n-1} |a^k|^2 \xi_l^k \rho^k \\
&= 4\pi^2( \sum_{k=1}^n \rho^k(\frac{1}{p_k}(\xi_l^k - \xi_l^{k-1})
-\frac{1}{p_{k+1}}(\xi_l^{k+1} - \xi_l^{k}) + |a^k|^2 \xi^k_l )
\label{eq:form}
\end{align}

If we equate each coefficient of $\rho^k$ in \eqref{eq:form}
to zero we can put the defining
condition for $\iota(\xi_i)$ into the following matrix form.
We let $\xi = ( \xi_l^k)$ be a matrix with rows labelled by $l$
and columns labelled by $k$. We denote by $X$ the diagonal
matrix whose $l$th diagonal entry is $|a^l|^2$. Finally
we denote by $P$ the following matrix:
$$
P = \begin{pmatrix}
 \frac{1}{p_1} + \frac{1}{p_2}     & -\frac{1}{p_2}
   & 0       & 0 &\dots& 0 \cr
              -\frac{1}{p_2} &\frac{1}{p_2} + \frac{1}{p_2}&
 -\frac{1}{p_3}& 0  & \dots & 0 \cr
             0 & - \frac{1}{p_3} &\frac{1}{p_3} + \frac{1}{p_4}&
 -\frac{1}{p_4}& \dots & 0 \cr
             \vdots & \vdots   & \vdots
   & \vdots         & \ddots & 0 \cr
             0 &   0 &     0 & 0&   -\frac{1}{p_n} &
                                       \frac{1}{p_{n-1}} +
 \frac{1}{p_{n}}\cr
\end{pmatrix}
$$
Then the condition satisfied by $\xi$ becomes the matrix equation:
$$
2\pi\xi(P+ X) =  X
$$
and the matrix we are trying to find, $K$, satisfies
$$
K = (1 - 2\pi \xi) X.
$$
It follows that
$$
K^{-1} = P^{-1} + X^{-1}.
$$
We conclude that the metric on $\N_c$ is of the form
\begin{equation}
g = \sum_{l,j = 1}^{n-1} \frac{1}{4\pi^2}(P^{-1} +
X^{-1})_{lj} \sum_{a=1}^3 dy_l^a dy_j^a +
\sum_{l,j=1}^{n-1}  (P^{-1} + X^{-1})^{-1}_{lj}
\alpha^l \alpha^j \label{eq:metricfinal}
\end{equation}
where $y_l = x^{l+1} - x^l = (1/2\pi)\mu^l$.

To finish we want to compare our result
\eqref{eq:metricfinal} to formula (7.5) in
\cite{LeeWeiPil1}. Except for rescalings the
 only question is to show that their matrix
$\mu_{ij}$ is the matrix $P^{-1}_{ij}$. To do this
we have to calculate $\mu_{ij}$ in the manner they suggest.
We reintroduce the center of mass co-ordinate $x_{\cm}$.
This means we replace $P^{-1}$ in \eqref{eq:metricfinal} by $\hat P^{-1}$
where
$$
\hat P^{-1} =
\begin{pmatrix}
p & 0 \\
0 & P^{-1}
\end{pmatrix}.
$$
Then we consider the effect of the co-ordinate change
from the co-ordinates $x^i$ to the co-odinates $(x_{\cm}, y^i)$.
This is the result of applying the linear transformation
$$
X = \begin{pmatrix}
\frac{p_1}{p}& \frac{p_2}{p}& \frac{p_3}{p} &\dots &
\frac{p_{n-1}}{p}& \frac{p_n}{p}\\
1 & -1 & 0 & \dots &0 & 0 \\
0 & 1 & -1 & \dots &0 & 0 \\
\vdots & \vdots & \vdots & \ddots& 0 & 0\\
0 & 0 & 0  & \dots & 1 & -1
\end{pmatrix}
$$
to the co-ordinates $x^i$. Hence the matrix of the
metric in terms of the co-ordinates $x^i$ is given by
$X^t P^{-1} X$. We leave it to the reader to check that
this is the diagonal matrix with entries $p_1, \dots, p_n$
which agrees with the definition of the constant term in
$M_{ii} $ in \cite{LeeWeiPil1} (their $m_i$ is our $p_i$).
So we conclude that the $\mu_{ij}$ in \cite{LeeWeiPil1}
is indeed $P^{-1}_{ij}$. The metric on $\N_c$ is
therefore the same asymptotically as the metric on
the monopole moduli space calculated in \cite{LeeWeiPil1}.

\bigskip

\noindent{\small \bf Acknowledgement} \\The author acknowledges
the support of the Australian Research Council.

\end{document}